\documentclass[aps,prb,groupedaddress,preprint]{revtex4-1}

\usepackage[english]{babel}
\usepackage[dvips]{graphicx}
\usepackage{indentfirst}
\usepackage{tabularx}

\begin{document}

\title
{Solidification of small para-H$_{\bf 2}$ clusters at zero temperature}
\author{E. Sola}
%\altaffiliation{Current address: Some other place, Othert\"own,
%Germany}
\affiliation
{Donostia International Physics Center, Donostia, Spain}
\author{J. Boronat}
\email{jordi.boronat@upc.edu}
\affiliation
{Department de F\'\i sica i Enginyeria Nuclear, Campus Nord B4-B5, E-08034
Barcelona, Spain}

%%%%%%%%%%%%%%%%%%%%%%%%%%%%%%%%%%%%%%%%%%%%%%%%%%%%%%%%%%%%%%%%%%%%%
%% The document title should be given as usual
%% A short title can be given as a *suggestion* for running headers.
%%%%%%%%%%%%%%%%%%%%%%%%%%%%%%%%%%%%%%%%%%%%%%%%%%%%%%%%%%%%%%%%%%%%%

%%%%%%%%%%%%%%%%%%%%%%%%%%%%%%%%%%%%%%%%%%%%%%%%%%%%%%%%%%%%%%%%%%%%%
%% The manuscript does not need to include \maketitle, which is
%% executed automatically.  The document should begin with an
%% abstract, if appropriate.  If one is given and should not be, the
%% contents will be gobbled.
%%%%%%%%%%%%%%%%%%%%%%%%%%%%%%%%%%%%%%%%%%%%%%%%%%%%%%%%%%%%%%%%%%%%%
\begin{abstract}
 We have determined the ground-state energies of para-H$_2$ clusters at
 zero temperature using the diffusion Monte Carlo method. The liquid or
 solid character of each
 cluster is investigated by restricting the phase through the use of proper
 importance sampling. Our results show inhomogeneous crystallization of 
 clusters, with alternating behavior between liquid and solid phases up to
  $N=55$. From there on, all clusters are solid. The ground-state energies in the range
 $N=13$--75 are established and the stable phase of each cluster is
 determined. In spite of the small differences observed between the energy
 of liquid and solid clusters, the corresponding density profiles are
 significantly different, feature that can help to solve ambiguities in the
 determination of the specific phase of H$_2$ clusters.
\end{abstract}

\maketitle
%%%%%%%%%%%%%%%%%%%%%%%%%%%%%%%%%%%%%%%%%%%%%%%%%%%%%%%%%%%%%%%%%%%%%
%% Start the main part of the manuscript here.
%%%%%%%%%%%%%%%%%%%%%%%%%%%%%%%%%%%%%%%%%%%%%%%%%%%%%%%%%%%%%%%%%%%%%
\section{Introduction}

Molecular para-hydrogen has been suggested from long time ago as the possible
second natural superfluid after liquid $^4$He.~\cite{ginzburg} However, the H$_2$-H$_2$
interaction is so deeply attractive that it crystallizes before arriving to
the superfluid transition temperature. The advantage of having half the
mass of $^4$He is therefore not enough to compensate the hydrogen strong
attraction and para-H$_2$ becomes solid at temperature $T=13.96$ K. Many
attempts to supercool liquid hydrogen down to the expected lambda
transition ($T_\lambda=1-2$ K) have been unfruitful, at least for the bulk
phase.\cite{clark} Partial success has only been achieved in confined geometries, mainly in
small pure liquid drops~\cite{tejeda} or bigger drops in a $^4$He
environment.~\cite{vilesov}

Para-H$_2$ clusters have been the object of many studies in the last
years~\cite{ceperley1,guardiola1,cuervo,bonin1}
due to the primary interest of determining \textit{i)} its liquid or solid character
and \textit{ii)} the dependence of superfluidity on the size and temperature of the
cluster. The first path integral Monte Carlo (PIMC) simulation of H$_2$
clusters, carried out by Sindzingre \textit{et al.},~\cite{ceperley1} showed that clusters
comprising up to a number of molecules $N \simeq 18$ are superfluid at
temperatures below $T=2$ K. The maximum number of molecules that shows
superfluid behavior has been enlarged up to $N \simeq 26$ in a
recent PIMC simulation where lower temperatures $T=0.5$ K have been
analyzed.~\cite{ceperley2} The results obtained in Ref. \cite{ceperley2} 
show evidence of superfluidity
mostly localized in the surface of the cluster, pointing to an inhomogeneous
structure with an inner solid core surrounded by a liquid \textit{skin}
that, at the 
lowest temperatures, is superfluid. This localized superfluidity has been
questioned in a recent PIMC calculation where it has been shown that superfluidity is
a global property of the cluster in spite of its significant spatial
structure.~\cite{bonin2} In the limit of zero temperature, the structure and energy of small H$_2$
clusters have been accurately studied using both diffusion Monte Carlo
(DMC)~\cite{guardiola1,guardiola2}
and path integral ground state (PIGS).~\cite{cuervo} Both at finite and 
zero temperature
the simulations show the presence of \textit{magic}-cluster sizes in
which the chemical potential shows a kink. These more stable $N$
configurations have also been observed experimentally in cryogenic free jet
expansions using Raman spectroscopy.~\cite{tejeda} 

In the present work, we address the question of the solidification of
para-H$_2$ clusters in the limit of zero temperature and as a function of
the number of molecules. Our aim is the screening of the more
stable ground-state structure by performing simulations where the phase
(liquid or solid) is kept fixed. To this end, we use the diffusion Monte
Carlo (DMC) method that solves stochastically the $N$-body Schr\"odinger
equation exactly for bosons, within some statistical errors. Differently
from previous studies using DMC, we have focused our attention on the
discrimination between liquid and solid phases as a function of the number
of molecules. This goal is achieved by carrying out parallel simulations
for liquid and solid configurations at each $N$, with special effort on the search
for optimal lattices on which to build trial wave functions for the
solid clusters. With the present study, we show which is the
energetically preferred phase at $T=0$ for each $N$, in the 
range $13 \leq N \leq 75$, and how the energy difference between the two
type of clusters changes with increasing $N$.

The rest of the paper is organized as follows. In the next Section, we
present our theoretical approach that relies on the DMC method. In Sec.
III, we report our results on the energetic and structure properties of the
para-H$_2$ clusters in the $N$ range analyzed. Finally, we end with the
summary and main conclusions in Sec. IV.

\section{Quantum Monte Carlo approach}

Our study of para-H$_2$ clusters relies on a purely microscopic approach
whose inputs are only the interparticle interaction and the mass. Our goal
is the study of these finite systems at zero temperature to deal with
their ground state. To this end, we use the DMC method which is able to
generate exact (within statistical uncertainty) information through guided
random walks. The starting point in the DMC method is the
$N$-body Schr\"{o}dinger equation, written in imaginary time
\begin{equation}
- \frac{\partial \Psi({\bf R},t)}{\partial t} = (H-E)\, \Psi({\bf R},t) \ ,
\label{dmc.eq1}
\end{equation}
with ${\bf R} \equiv ({\bf r}_1,\ldots,{\bf r}_N)$, a $3N$-dimensional
vector ({\em walker}), and $t$ the imaginary time measured in
units of $\hbar$. The time-dependent 
wave function of the system $\Psi({\bf R},t)$ can be expanded in terms of a
complete set of eigenfunctions $\phi_i({\bf R})$ of the Hamiltonian,
\begin{equation}
\Psi({\bf R},t)=\sum_{n}c_n \, \exp \left[\, -(E_i-E)t \, \right]\,
\phi_i({\bf R})\ ,
\label{dmc.eq1b}
\end{equation}
where $E_i$ is the eigenvalue associated to $\phi_i({\bf R})$. 
Consequently, the
asymptotic solution of \ref{dmc.eq1}, for any value $E$ close to
the energy of the ground state and for long times ($t \rightarrow
\infty$), gives $\phi_0({\bf R})$, provided that there is a nonzero
overlap between $\Psi({\bf R},t=0)$ and the ground-state wave function
$\phi_0({\bf R})$.

A direct Monte Carlo implementation of \ref{dmc.eq1} is hardly
able to work efficiently, especially when the interatomic potential
contains a hard core. This is solved by using importance sampling. 
The importance sampling technique is
a general concept in Monte Carlo and is one of the best
methods to reduce the variance of any MC calculation. The importance
sampling method, applied to \ref{dmc.eq1}, consists in
rewriting the Schr\"{o}dinger equation in terms of the wave function
\begin{equation}
f({\bf R},t)\equiv \psi({\bf R})\,\Psi({\bf R},t)\ ,
\label{dmc.eq2}
\end{equation}
$\psi({\bf R})$ being a time-independent variational wave function that describes
approximately the ground state of the system. Considering a Hamiltonian of the form
\begin{equation}
H=-\frac{\hbar^2}{2\,m} \, {\mbox{\boldmath $\nabla$}}^2_{{\bf R}} 
+ V({\bf R})\ ,
\label{dmc.eq3}
\end{equation}
\ref{dmc.eq1} turns out to be
\begin{equation}
-\frac{\partial f({\bf R},t)}{\partial t}  =  -D\, 
{\mbox{\boldmath $\nabla$}}^2_{{\bf R}}
f({\bf R},t)+D\,{\mbox{\boldmath $\nabla$}}_{{\bf R}} \left( {\bf F}({\bf R})
\,f({\bf R},t)\,
\right)+\left(E_L({\bf R})-E \right)\,f({\bf R},t)  \ , 
\label{dmc.eq4}
\end{equation}
where $D=\hbar^2 /(2m)$, $E_L({\bf R})=\psi({\bf R})^{-1} H \psi({\bf R})$
is the local energy, and
\begin{equation}
{\bf F}({\bf R}) = 2\, \psi({\bf R})^{-1} 
{\mbox{\boldmath $\nabla$}}_{{\bf R}} \psi({\bf R})
\label{dmc.eq5}
\end{equation}
is called drift or quantum force. ${\bf F}({\bf R})$ acts as an external force
which guides the diffusion process, involved by the first term in 
\ref{dmc.eq4}, to regions where $\psi({\bf R})$ is large.

The r.h.s. of \ref{dmc.eq4} may be written as the action of three
operators $A_i$ acting on the wave function $f({\bf R},t)$,
\begin{equation}
-\frac{\partial f({\bf R},t)}{\partial t} = (A_1+A_2+A_3)\, 
f({\bf R},t) \equiv A\, f({\bf R},t)
\label{dmc.eq4p}
\end{equation}
The three terms $A_i$ may be interpreted by similarity with classical
differential equations. The first one, $A_1$, corresponds to a free
diffusion with a a diffusion coefficient $D$; $A_2$ acts as a driving force
due to an external potential, and  finally $A_3$ looks like a birth/death
term.
 In Monte Carlo, the Schr\"odinger equation \ref{dmc.eq4p}  is best
suited when it is written in an integral form by introducing the Green
function $G({\bf R}^{\prime},{\bf R},t)$, which gives the
transition probability  from an initial state ${\bf R}$ to a final one 
${\bf R}^{\prime}$ during a time $t$, 
\begin{equation}
     f({\bf R}^{\prime},t+\Delta t) =\int G({\bf R}^{\prime},{\bf R},
\Delta t)\, f({\bf R},t)\, d{\bf R} \ .
\label{dmc.eq6}
\end{equation}
More explicitly, the Green function is given in terms of the operator
$A=A_1+A_2+A_3$ by
\begin{equation}
    G({\bf R}^{\prime},{\bf R}, \Delta t) =  
    \left \langle\,
{\bf R}^{\prime}\, | \, \exp(-A \Delta t)\, |\, {\bf R}\, \right \rangle.
\label{dmc.eq7}
\end{equation}

DMC algorithms rely on reasonable approximations of $G({\bf R}^{\prime},{\bf
R},\Delta t)$ for small values of the time-step $\Delta t$. 
We work with a second-order expansion of the exponential [\ref{dmc.eq7}] to
reduce the time-step dependence.~\cite{boronat} Once a short-time approximation is
assumed, \ref{dmc.eq7}
is iterated repeatedly until reaching the asymptotic regime
$f({\bf R},t \rightarrow \infty)$, a limit in which one is effectively
sampling the ground state.  

Para-H$_2$ is a boson particle with total spin 0 and with rotational
ground-state state $J=0$. It is well described by a merely radial interaction
due to its high-degree sphericity. Among the different interatomic
potentials proposed for describing the H$_2$-H$_2$ intermolecular
interaction,
we have chosen the Silvera-Goldman potential~\cite{silvera} due to its proved accuracy and
its dominant use in microscopic calculations as the present one. The well
depth of the molecular hydrogen interaction is $\sim -37$ K, a factor of four
larger than in helium, making the ground-state phase of bulk at zero
temperature be an hcp crystal in spite of H$_2$ mass being half the $^4$He
one.

The trial wave function used for importance sampling is written as the
product of one- and two-body correlation factors,
\begin{equation}
\psi({\bf R}) = \prod_{1=i<j}^{N} e^{u(r_{ij})} \, \prod_{i=1}^{N}
e^{v(r_{iI})} \ ,
\label{eq.jastrow}
\end{equation}
with
\begin{eqnarray}
u(r) & = & -\frac{1}{2} \left( \frac{b}{r} \right)^5 - \frac{\beta r}{N}
\label{eq.ujas} \\
v(r) & = & - \alpha r^2 \ .
\label{eq.vjas}
\end{eqnarray}
The two-body term $u(r)$ accounts for the correlations induced by the
potential $V(r)$ and also for the finite size of the system that implies 
the wave function approaches zero when the distance is of the order of the
cluster size. The one-body term $v(r)$ is only used for solid
clusters and localizes particles around the preferred sites
(capital indexes in \ref{eq.jastrow}). 
This model wave function for the solid, with the one-body Gaussian terms, is the
well-know Nosanow-Jastrow (NJ) wave function that has been widely used in the
study of bulk quantum solids. The NJ wave function is not symmetric under
the exchange of particles but the influence in the energy of 
symmetrization is known to be of the order of miliKelvin~\cite{bernu} and
therefore indistinguishable within our statistical errors. Recently, it has been
proposed a symmetrized NJ wave function to study superfluidity in bulk
solid $^4$He and proved that the change in energy due to Bose symmetry on
top of the NJ model is imperceptible, within the numerical resolution of
quantum Monte Carlo.~\cite{cazorla} It is worth noticing that the effect of symmetrization 
in the structure of para-H$_2$ clusters  at temperature $T=1.5$ K 
has been recently studied by Warnecke \textit{et al.};~\cite{warnecke} the results obtained
show a very small influence of Bose statistics on the density profiles,
the energy differences being not reported. 

The parameters entering in Eq. (11) and Eq. (12) have been optimized using
the variational Monte Carlo (VMC) method. For liquid configurations  the
optimal parameters  are: $b=3.58$ \AA, $\beta=2.79$ \AA$^{-1}$, and
$\alpha=0$; for solid ones: $b=3.32$ \AA, $\beta=0$, and $\alpha=0.521$
\AA$^{-2}$.  These are the VMC optimal parameters for $N=40$ but  the
dependence of these parameters with $N$ is tiny: for $N=13$,  $\beta=2.30$
\AA$^{-1}$, $\alpha=0.360$ \AA$^{-2}$, and $b$ is the same. In DMC we have
neglected this slight $N$ dependence because the results are insensitive to
it.  On the other hand, two technical issues related to the implementation
of the DMC method, i.e., the time-step dependence and the number of
walkers, have been accurately analyzed to reduce any systematic bias to the
level of the statistical noise.

The simulation of solid clusters with the NJ wave
function we are using requires of a set of lattice points. 
In the case of bulk solids, the number of
possible lattices is few, well known and easy to characterize
geometrically. When dealing with finite systems, the issue of the best
geometrical arrangement is somehow universal for the smallest structures,
where Mackay polyhedra are the preferred structures, but it becomes more complex when
the number of particle increases. Moreover, the determination of the
optimal solid patterns are normally obtained using classical physics and
the influence on those of quantum delocalization is less known. Our
strategy for this optimization has been the use of both simulated annealing
(SA) and ab initio random search (method for finding structures where the total classical force
felt by any particle is approximately zero). The structures of minimum
energy predicted for both approaches essentially coincide for $N \leq 30$,
but for larger $N$ values the SA method has proven to be better. In the SA
approach, we have used both an exponential annealing schedule and another
with constant thermodynamic speed.~\cite{McKeown} The search for any $N$ has been carried
out by starting from 20--40 random configurations and selecting the best
ones for a posterior quantum simulation. In some cases, we have performed
additional optimizations using long classical Monte Carlo simulations.
Finally, we used the best classical solutions as initial setup for the
quantum simulations and reported in all cases the lowest energy achieved.
It is worth mentioning that we introduced a scaling factor $f_{\rm s}$ 
in the quantum simulation, 
\begin{equation}
{\bf r}_i^{\rm s} = ( {\bf r}_i - {\bf r}_{\rm CM} ) f_{\rm s} \ ,
\label{scaling}
\end{equation}
with ${\bf r}_{\rm CM}$ the center of mass of the cluster and
${\bf r}_i$ ($i=1,\ldots,N$) the SA points,
to allow for possible spatial expansions or contractions of the
classic lattice points but the optimal energy was always the corresponding
to the classic solutions ($f_{\rm s}=1$).

\section{Results}            

\begin{figure}[tb]
\begin{center}
\includegraphics[angle=0,width=0.7\textwidth]{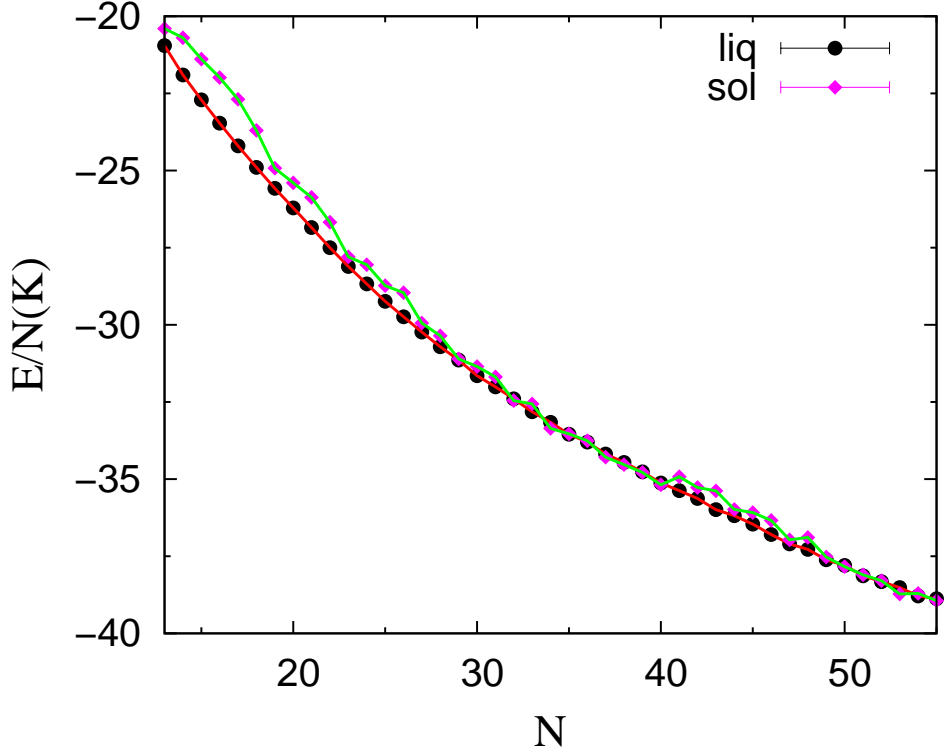}
\caption{(Color online) Energy per particle of H$_2$ clusters as a function
of $N$. Circles and diamonds stand for liquid and solid phases,
respectively. The lines are guides to the eye. Error bars are smaller than
the size of the symbols }
\label{Fig:1}
\end{center}
\end{figure}

We have calculated the energy and structure properties of H$_2$ clusters in
the range $13 \leq N \leq 75$ using the DMC method discussed in the
preceding Section. At each $N$, we have used both the liquid and solid
trial wave functions introduced in Sec. II. In \ref{Fig:1}, we show
the results for the energy per particle obtained for both phases and for
clusters up to $N=55$ molecules. The
difference between both configurations is small in all the $N$ range
studied pointing to a highly correlated liquid, even for the lightest
clusters. The energy per particle in absolute value increases with $N$
for both phases but solid clusters have energies with a kind of
zigzag dependence whereas the liquid ones follow a smoother law. The irregular behavior
of the energy of solid clusters is consequence of the appearance of magic
numbers, with geometrically more compact structures that make them more stable.

\begin{figure}[tb]
\begin{center}
\includegraphics[angle=0,width=0.7\textwidth]{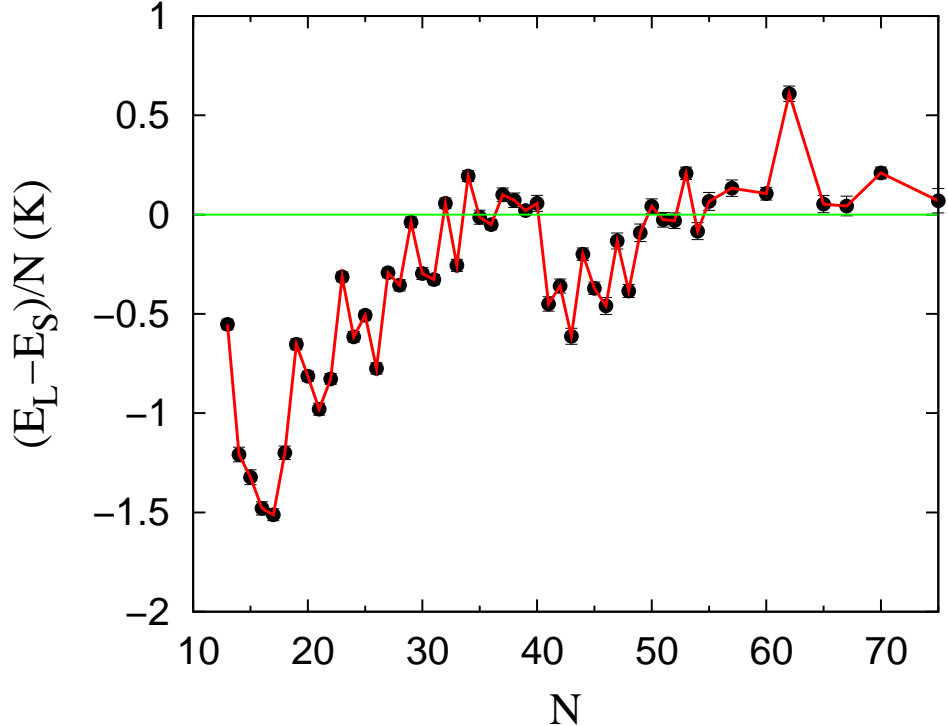}
\caption{(Color online) Difference between the energies per particle of
liquid and solid clusters ($(E_L-E_S)/N$) as a function of $N$.}
\label{Fig:2}
\end{center}
\end{figure}

The difference between the energies of liquid and solid clusters is shown
explicitly in \ref{Fig:2}. Starting from $N=13$, the energy of liquid
clusters is clearly preferred but the difference $(E_L-E_S)/N$  decreases
when $N$ increases. According to our results, the first cluster that is
solid in its ground state is the one with $N=32$. From $N=33$ to 40, there
are some solid clusters but for $N=41$ and the following ones there is
again a liquid regime. Arriving to $N=55$, one enters definitively in a
preferred solid phase that persists up to $N=75$ which is the largest
cluster here analyzed. The more intriguing aspect of our results is the
existence of a liquid stability \textit{island} for $N=41$--50 between the
initial solid regime and the second one, that we think is the stable one for
$N > 55$. It can be argued that the nonuniform crystallization of H$_2$
clusters that we have observed can be due to our model of solid
clusters. We can not discard completely this argument but we have
performed a rather exhaustive search of lattice points, as commented in the
preceding Section, and the results remain unchanged.

\begin{figure}[tb]
\begin{center}
\includegraphics[angle=0,width=0.7\textwidth]{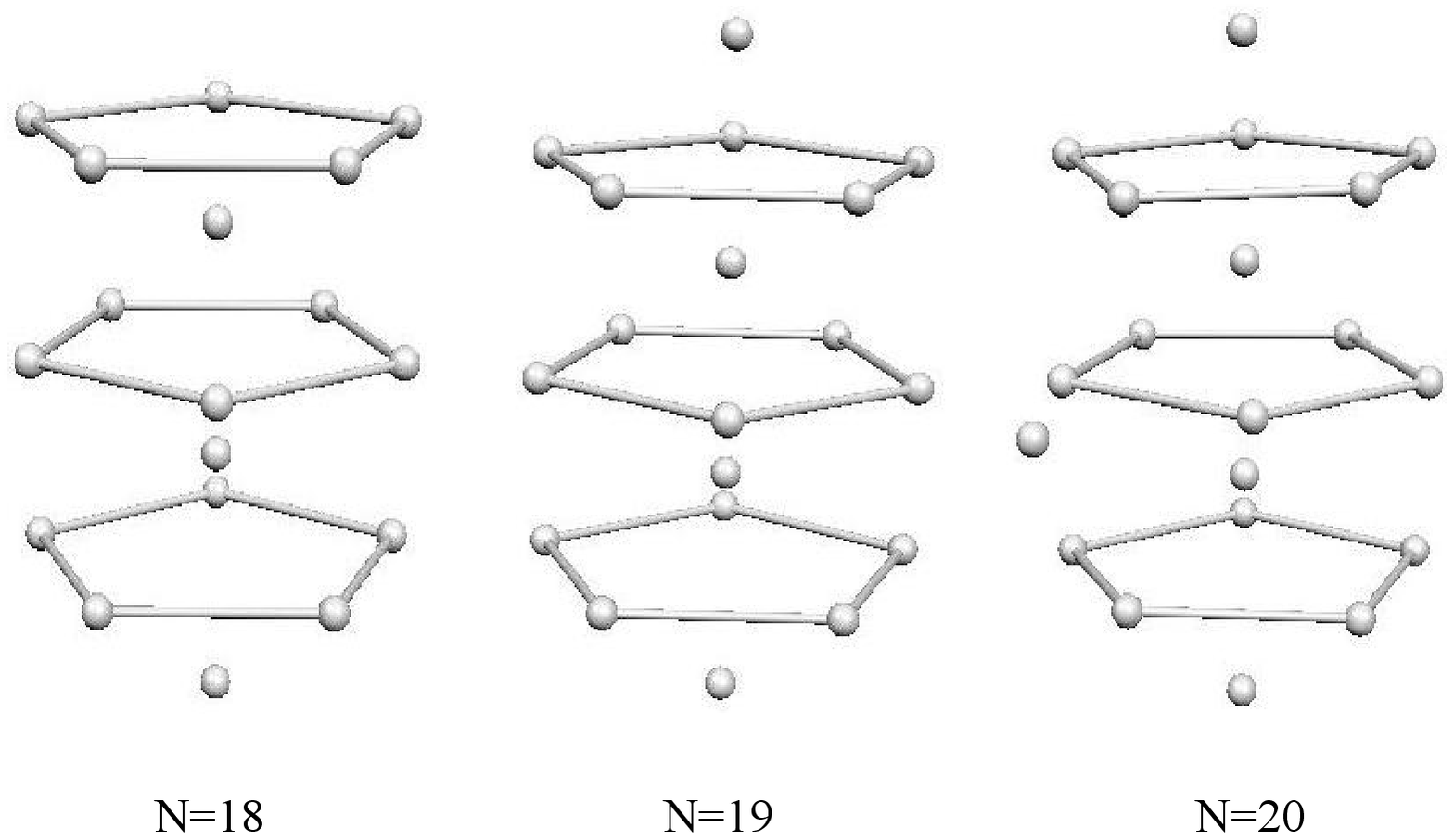}
\caption{(Color online) Optimal distribution of lattice points for solid
clusters with $N=18$, 19, and 20.}
\label{Fig:3}
\end{center}
\end{figure}

An important part of our simulation has been the search of optimal
structures for solid clusters. The most useful tool has been the simulated
annealing algorithm that, in spite of being completely classical, has proven
to be able to generate the best configurations. A simple check for that
has been the optimization of a scaling parameter $f_{\rm s}$ that we introduced to
make possible contractions or expansions of the classical solution.
Systematically, the optimal solution has been $f_{\rm s}=1$. In \ref{Fig:3},
the best lattice points for solid clusters with $N=18$, 19, and 20 are
shown. We have joined with lines the planar structures to get a better
visualization of the clusters. The case $N=19$ corresponds to a well-know
magic number and it is in fact a structure that we identify in a lot of
clusters. As one can see in \ref{Fig:3}, in the clusters $N=18$ and
$N=20$ the one-defect and one-excess structures, with
respect to the geometrically perfect $N=19$ lattice, are clearly observed.

\begin{figure}[tb]
\begin{center}
\includegraphics[angle=0,width=0.7\textwidth]{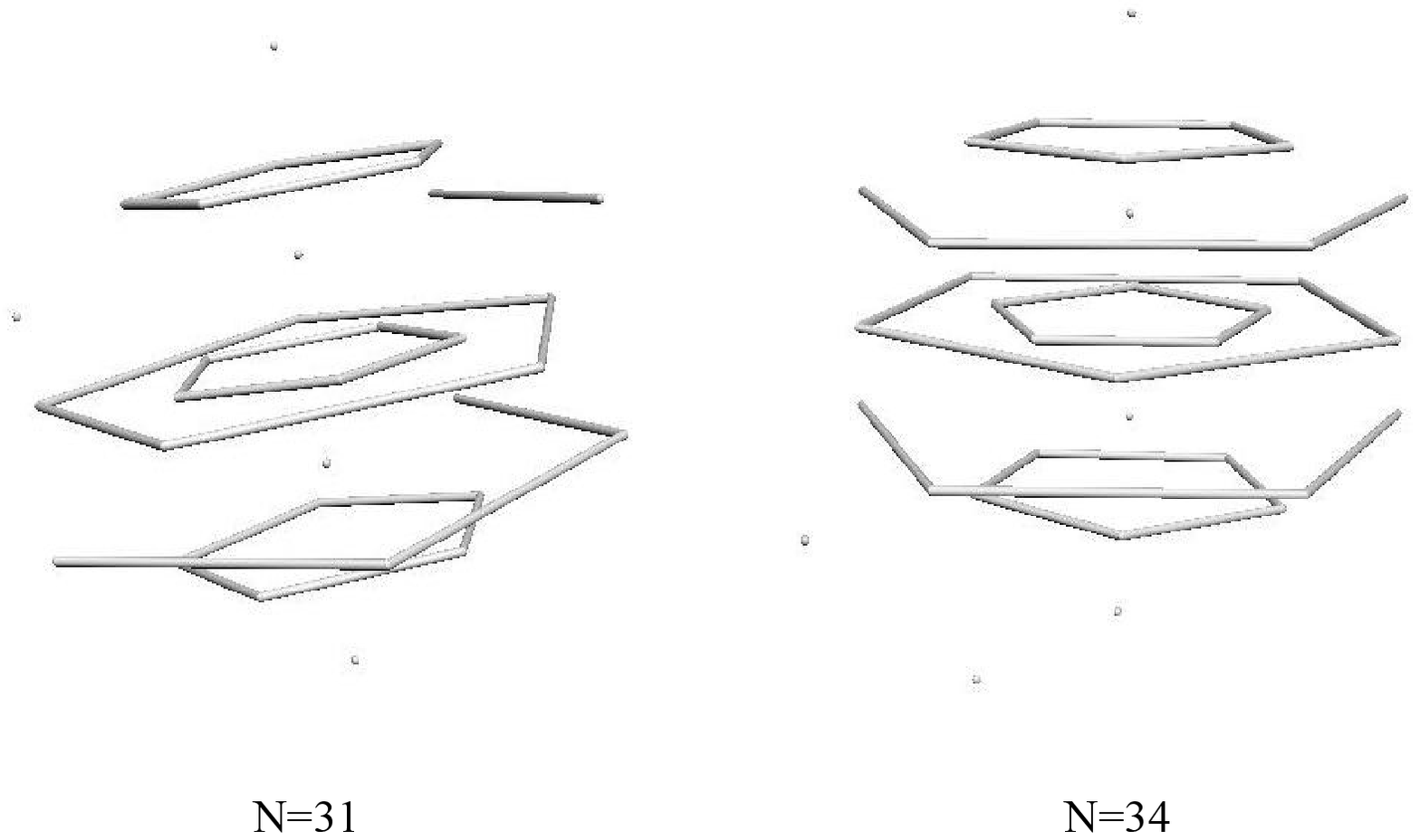}
\caption{(Color online) Optimal distribution of lattice points for solid
clusters with $N=31$ and 34.}
\label{Fig:4}
\end{center}
\end{figure}

The effort of achieving optimal structures increases significantly with
$N$,
so a lot of SA simulations with different initial configurations have been
carried out. Our best configurations coincide with reported ones for
Lennard-Jones interactions up to $N \simeq 30$.~\cite{web} For larger $N$ values, our
results differ significantly from the published ones~\cite{web} and we get
systematically better energies with our structures. In \ref{Fig:4}, we
report our optimal lattice points for clusters with $N=31$ and $N=34$,
i.e., beyond the regime where our predictions are compatible with the ones
from Ref. \cite{web}. As commented before, it is quite remarkable that the inner
structure of these two clusters is the same and coincides exactly with the
magic structure of $N=19$. Around this well-defined structure it starts to
appear a second pentagonal shell which is only complete in the center and
concentric with the central pentagon of the $N=19$ cluster.

In Table 1, we report the ground-state energies for each $N$ and the
corresponding phase.  Our results for the liquid clusters agree with the
previous results of Ref.  \cite{guardiola2} obtained also using the DMC method and the same
interaction potential. The present results for the solid clusters are new and never
calculated before. The energies reported in Table 1 lead to persistence of liquid 
character up
to relatively large $N$ values, in particular to larger values than the
ones reported in previous PIMC estimations at finite
temperature.~\cite{ceperley2,bonin2}
Nevertheless, the difference in energy between the two phases remains small
even for the largest clusters studied. We made several attempts at
simulating clusters with an inner solid core and a liquid surface but we
were not able to get any improvement of the energy with respect to the
optimal values reported in the table.

\begin{table}
\caption{Energy per particle of the ground state of para-H$_2$ clusters as
a function of $\bf{N}$. The phase of the cluster is labeled L and S for liquid
and solid, respectively. Figures in parenthesis are the statistical errors.}
\begin{tabular}{ccc|ccc} 
\hline
    $N$    &        $E/N$ (K) &  Phase  &  $N$    &   $E/N$ (K) &  Phase  \\  
     \hline
     13    &     $-20.952(16)$       &   L &  38         &     $-34.531(16)$     &   S      \\ 
     14    &     $-21.904(14)$       &	 L &  39	 &     $-34.785(17)$	   &   S       \\   
     15    &     $-22.709(11)$       &	 L &  40	 &     $-35.184(17)$	   &   S     \\
     16    &     $-23.466(12)$       &	 L &  41	 &     $-35.376(27)$	   &   L     \\
     17    &     $-24.199(16)$       &	 L &  42	 &     $-35.630(27)$	   &   L    \\
     18    &     $-24.901(17)$       &	 L &  43	 &     $-35.993(35)$	   &   L     \\
     19    &     $-25.576(17)$       & 	 L &  44	 &     $-36.184(25)$	   &   L      \\
     20    &     $-26.215(14)$       & 	 L &  45	 &     $-36.459(27)$	   &   L     \\
     21    &     $-26.847(16)$       &	 L &  46	 &     $-36.796(38)$	   &   L     \\
     22    &     $-27.500(18)$       &	 L &  47	 &     $-37.099(31)$	   &   L     \\
     23    &     $-28.111(12)$       &	 L &  48	 &     $-37.278(28)$	   &   L       \\
     24    &     $-28.671(18)$       &	 L &  49	 &     $-37.615(38)$	   &   L    \\
     25    &     $-29.237(19)$       &	 L &  50	 &     $-37.837(24)$	   &   S    \\
     26    &     $-29.737(17)$       &	 L &  51	 &     $-38.131(32)$	   &   L    \\
     27    &     $-30.235(14)$       &	 L &  52	 &     $-38.326(33)$	   &   L    \\
     28    &     $-30.710(22)$       &	 L &  53	 &     $-38.726(11)$	   &   S    \\
     29    &     $-31.141(22)$       &	 L &  54	 &     $-38.787(39)$	   &   L    \\
     30    &     $-31.646(24)$       &	 L &  55	 &     $-38.942(23)$	   &   S     \\
     31    &     $-32.013(22)$       &	 L &  57	 &     $-39.382(24)$	   &   S    \\
     32    &     $-32.454(15)$       &	 S &  60	 &     $-39.820(19)$	   &   S    \\
     33    &     $-32.818(26)$       &	 L &  62	 &     $-40.629(19)$	   &   S     \\
     34    &     $-33.357(15)$       &	 S &  65	 &     $-40.543(19)$	   &   S     \\
     35    &     $-33.547(27)$       &	 L &  67	 &     $-40.882(19)$	   &   S     \\
     36    &     $-33.804(19)$       &	 L &  70	 &     $-41.409(17)$	   &   S    \\
     37    &     $-34.287(18)$       &	 S &  75	 &     $-41.711(24)$	   &   S   \\
 \hline
\end{tabular}

\end{table}

\begin{figure}[tb]
\begin{center}
\includegraphics[angle=0,width=0.7\textwidth]{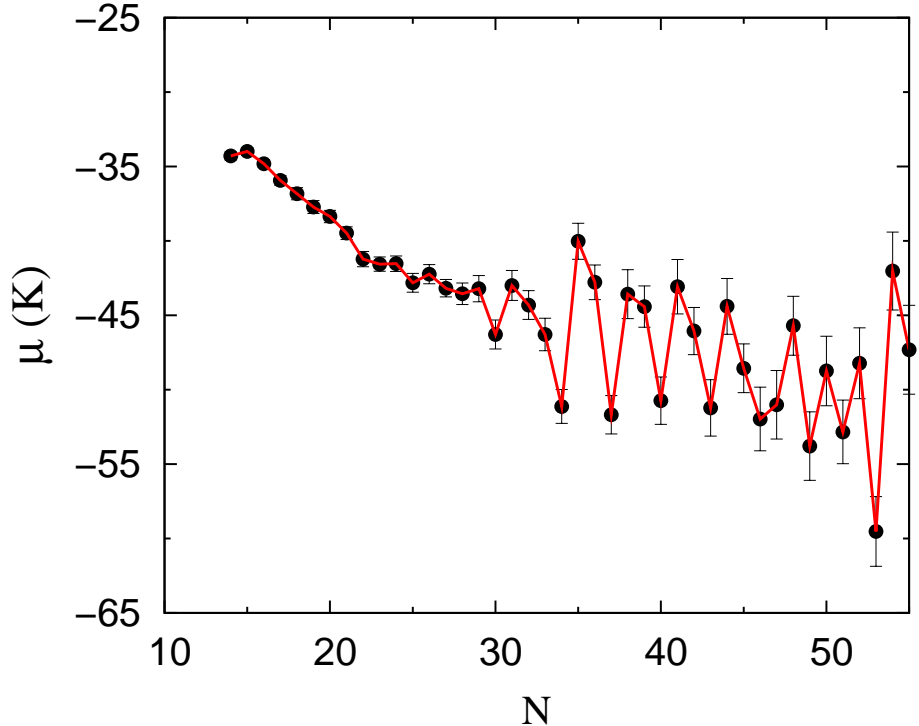}
\caption{(Color online) Chemical potential of H$_2$ clusters as a function
of $N$.}
\label{Fig:5}
\end{center}
\end{figure}

The possibility of magic clusters, with higher probability of being
experimentally observed due to its enhanced stability, has deserved the
attention of all the theoretical and experimental works in this subject. If
such structures exist, one can see its signature in energy differences,
mainly in the chemical potential defined as
\begin{equation}
\mu(N) = E(N) -E(N-1) \ .
\label{potquim}
\end{equation}
In \ref{Fig:5}, we show the results for the chemical potential
[\ref{potquim}] obtained from the DMC ground-state energies reported in Table
1. In the liquid regime (small clusters), $\mu(N)$ is quite a smooth
function with small local minima for $N=23$, 25, and 30. For $N > 30$, a
clear zigzag structure is observed with some minima that correspond to
liquid clusters and other that are solid. Therefore, the chemical potential
we obtain is no more a smooth function as derived in previous DMC
estimations where only liquid clusters where
considered.~\cite{guardiola1,guardiola2} Our results show a
more complex structure with some alternating phases and with a lot of local
minima, mainly when entering in the solid phase.

\begin{figure}[tb]
\begin{center}
\includegraphics[angle=0,width=0.7\textwidth]{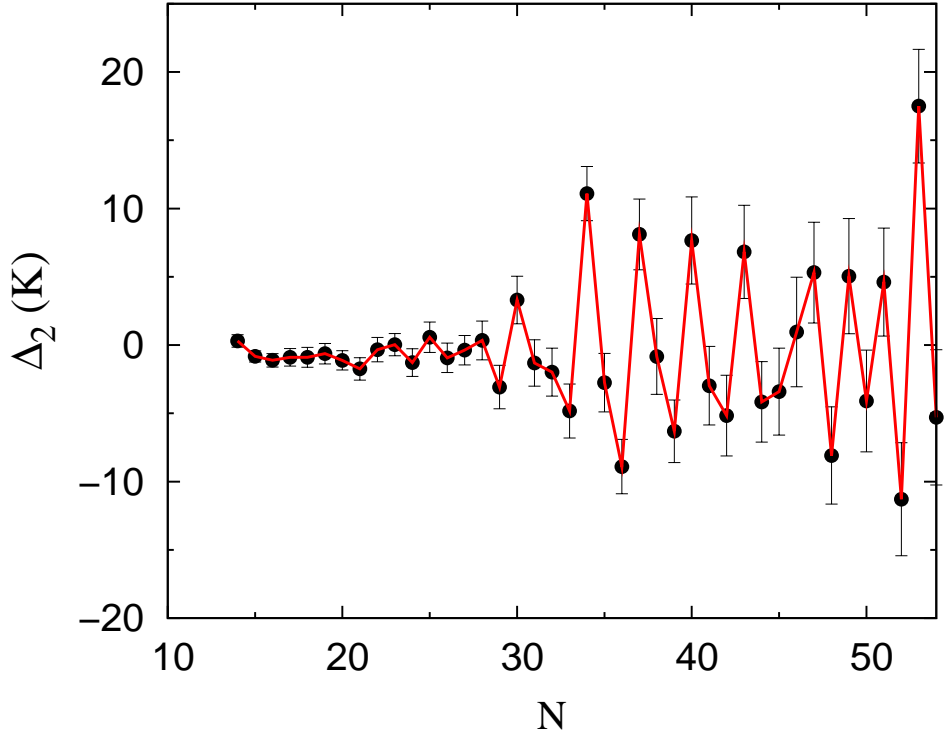}
\caption{(Color online) Second energy difference $\Delta_2$ 
of H$_2$ clusters as a function of $N$.}
\label{Fig:5b}
\end{center}
\end{figure}

The issue of the existence of magic clusters can be better analyzed by
calculating the second energy difference,
\begin{equation}
\Delta_2(N) = E(N+1) - 2 E(N) + E(N-1) \ .
\label{delta2}
\end{equation}
DMC results for $\Delta_2(N)$ are shown in Fig. \ref{Fig:5b}. The observed
pattern emphasizes the behavior of $\mu(N)$, with a quite smooth behavior
up to $N \sim 30$ and sharp peaks beyond this value. By looking at the
position of the maxima, we observe magic structures for $N=13$, 19, 23, 25,
28, 30, 34, 37, 40, 43, 47, 49, 51, and 53. Some of these values have been
recently found in PIMC simulations at finite
temperatures.~\cite{navarro,mezzacapo}

\begin{figure}[tb]
\begin{center}
\includegraphics[angle=0,width=0.7\textwidth]{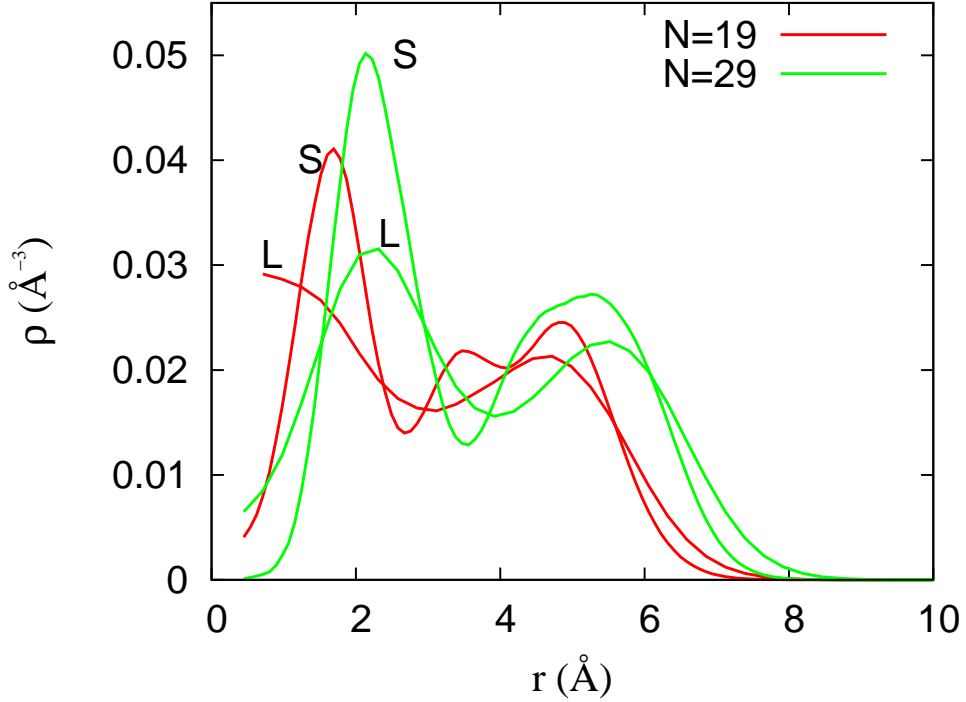}
\caption{(Color online) Density profiles of H$_2$ clusters with $N=19$ and
29. Labels L and S stand for liquid and solid clusters, respectively. }
\label{Fig:6}
\end{center}
\end{figure}

DMC simulations serve also to calculate structure properties of the
clusters, as for instance, the density profiles. In \ref{Fig:6}, we
show the density profiles of clusters with $N=19$ and $N=29$ using both the
liquid and solid trial wave functions. As one can see, the \textit{size} of
the cluster is very similar for liquid and solid clusters with the same $N$
but the inner structure is rather different. This is particularly clear for
$N=19$ where the central density of the liquid is large whereas for the
solid is zero. This central zero density for the solid cluster can be well understood by
inspection of the lattice points of the $N=19$ cluster reported in
\ref{Fig:3}. When $N$ increases, the differences between liquid and
solid clusters diminish but are still clear for the case $N=29$ reported in
\ref{Fig:6}. It is worth noticing that the density profiles of the
liquid clusters present a two-shell
effect~\cite{guardiola1,cuervo,ceperley1,bonin1} due to the strong intermolecular
H$_2$-H$_2$ attraction that is never observed in liquid $^4$He clusters.

\begin{figure}[tb]
\begin{center}
\includegraphics[angle=0,width=0.7\textwidth]{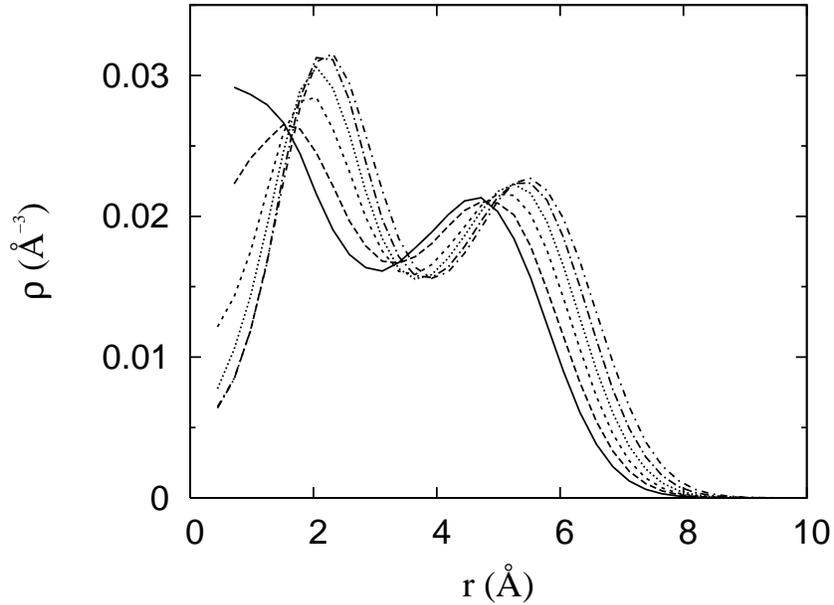}
\caption{Density profiles of liquid H$_2$ clusters with $N=19$, 21, 23, 25,
27, and 29. The line for each $N$ can be identified looking at the radius
of the cluster that increases monotonically with $N$. }
\label{Fig:7}
\end{center}
\end{figure}

\begin{figure}[tb]
\begin{center}
\includegraphics[angle=0,width=0.7\textwidth]{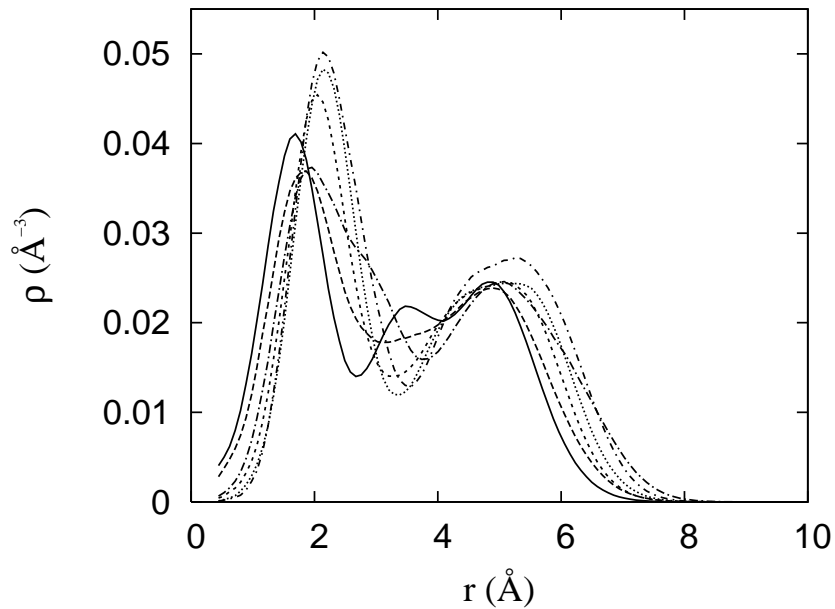}
\caption{Density profiles of solid H$_2$ clusters with $N=19$, 21, 23, 25,
27, and 29. The line for each $N$ is of the same type than in Fig. 7.	} 
\label{Fig:8}
\end{center}
\end{figure}

We report the evolution of the density profiles of liquid and solid
clusters with $N$ in \ref{Fig:7} and \ref{Fig:8}, respectively.
Increasing $N$ in the liquid clusters produces a continuous change in the
density profile: the central density decreases to zero and a two-shell
structure, that moves progressively to larger $r$, clearly appears.  
In the case of solid clusters (\ref{Fig:8}), the evolution with $N$ is
not so smooth, mainly for the lowest $N$ values. Different from the liquid
clusters, solid ones show always empty density in the center of the cluster
and additional structure between the two shells that tend to dissappear when
$N$ becomes larger. Comparing density profiles for the same $N$ and
different phase, one concludes that functions $\rho(r)$ are in all cases different
enough to be unambiguously discerned in spite of the fact that the
difference between their binding energies is less that 1 K per
particle.

\section{Conclusions}

H$_2$ clusters in the range $N=13$--75 have been microscopically
characterized at zero temperature using the DMC method. Controlling the
phase of the cluster by using different models for trial wave functions
used for importance sampling we have been able to determine the
ground-state stable phase for each $N$ in the range analyzed. Our results
point to a nonuniform crystallization of the H$_2$ clusters, with some
alternating behavior between the two phases depending on the particular $N$ value. For
clusters with $N \geq 55$ the stable phase is the solid one. The structure
of the clusters, as shown in their density profiles, is significantly
different for liquid and solid clusters with the same $N$, even when the
difference in energy between both is really tiny. Therefore, the shape of
the density profiles can help to identify the nature of these clusters both
in experiment and in finite-temperature simulations.

%%%%%%%%%%%%%%%%%%%%%%%%%%%%%%%%%%%%%%%%%%%%%%%%%%%%%%%%%%%%%%%%%%%%%
%% The "Acknowledgement" section can be given in all manuscript
%% classes.  Rather than use \section, an appropriate macro is
%% provided that will always work.
%%%%%%%%%%%%%%%%%%%%%%%%%%%%%%%%%%%%%%%%%%%%%%%%%%%%%%%%%%%%%%%%%%%%%
\section*{Acknowledgements}
We acknowledge partial financial support from DGI (Spain) Grant No.
FIS2008-04403 and Generalitat de Catalunya Grant No. 2009SGR-1003.

%%%%%%%%%%%%%%%%%%%%%%%%%%%%%%%%%%%%%%%%%%%%%%%%%%%%%%%%%%%%%%%%%%%%%
%% The same is true for Supporting Information, which should use the
%% \suppinfo macro.
%%%%%%%%%%%%%%%%%%%%%%%%%%%%%%%%%%%%%%%%%%%%%%%%%%%%%%%%%%%%%%%%%%%%%

%%%%%%%%%%%%%%%%%%%%%%%%%%%%%%%%%%%%%%%%%%%%%%%%%%%%%%%%%%%%%%%%%%%%%
%% The appropriate \bibliography command should be placed here.
%% Notice that the class file automatically sets \bibliographystyle
%% and also names the section correctly.
%%%%%%%%%%%%%%%%%%%%%%%%%%%%%%%%%%%%%%%%%%%%%%%%%%%%%%%%%%%%%%%%%%%%%
%\bibliography{achemso}

\end{document}